# Piezoelectric Voltage Coupled Reentrant Cavity Resonator

N. C. Carvalho,[1,2a)] Y. Fan,[1,2] J-M. Le Floch,[1,2] and M. E. Tobar[1,2]

[1]School of Physics, The University of Western Australia, 35 Stirling Hwy, 6009 Crawley, Western Australia, Australia
[2]ARC Centre of Excellence, Engineered Quantum Systems (EQuS), The University of Western Australia, 35 Stirling Hwy, 6009 Crawley, Western Australia, Australia

A piezoelectric voltage coupled microwave reentrant cavity has been developed. The central cavity post is bonded to a piezoelectric actuator allowing the voltage control of small post displacements over a high dynamic range. We show that such a cavity can be implemented as a voltage tunable resonator, a transducer for exciting and measuring mechanical modes of the structure and a transducer for measuring comparative sensitivity of the piezoelectric material. Experiments were conducted at room and cryogenic temperatures with results verified using Finite Element software.

## I. INTRODUCTION

Microwave reentrant cavities are typically made up of a right cylinder with a central post and have been widely investigated[1-13]. The post acts as a capacitor, creating a very intense electric field at the central region of the cavity, in the gap between the post and the adjacent wall. The magnetic field, on the other hand, circulates the post, composing the inductive region of the cavity[14]. Adjusting the gap spacing between the post and the cavity's wall allows this cavity to be employed as a highly tunable or displacement sensitive microwave resonator, making such a device very useful for a wide range of applications, such as solid state microwave oscillators[15], particle accelerators[16], dielectric characterization[17], electron spin resonance spectroscopy[18], electromechanical transducers for gravitational wave detectors[12,19] and test of fundamental physics experiments[20,21], and more.

Microwave cavities coupled to piezoelectric (PZT) crystals have also been developed for several purposes in the last decades, patents[22,23] from 1969 and 1978 reveal that methods to tune cavity resonant frequencies through PZT devices have long been explored. Since then, more elaborated structures than the former cavities with tunable walls are being created. Examples of how PZT devices still attract great scientific interest for resonators design can be found on recent works, as the micro-strip resonator at reference[24] or the Whispering Gallery resonator described by reference[25]. This work, despite these, presents a completely novel highly tunable voltage coupled piezoelectric reentrant cavity, where the PZT actuator is used to tune the resonance frequency through the cavity post not the cavity wall and it takes benefit of the unique features of reentrant cavities.

We have already developed a highly tunable cavity with a screw mechanism to mechanically tune the gap[26]. However, the drawback of such a device is the way to control the tuning mechanism, which has to be adjusted by hand and could not be easily controlled electronically without the development of a more complex structure; such has been implemented previously to tune sapphire Whispering Gallery mode resonators using stepper motors at cryogenic temperatures[27]. Now, differently, we couple a PZT ceramic actuator to the cavity post, which allows finer control using a DC voltage source. Such a device is more versatile as it can be more easily adapted to work at cryogenic temperatures. In our case, we have implemented the device in a 4 K cryogen-free pulse-tube cryocooler, which will be useful for a range of low temperature physics experiments.

This device demonstrated notable performance as a very sensitive tunable resonator, allowing us to idealize and test innovative applications for such a mechanism. As an electromechanical sensor, it can be used as tool for investigating electromechanical systems, exciting and measuring mechanical modes, and for PZT characterization at room and cryogenic temperatures.

## II. VOLTAGE TUNABLE RESONANT CAVITY EXPERIMENT

The PZT actuator was attached to the post in such a way that applying a DC voltage to the piezoelectric material induced a longitudinal displacement of the cavity post, and thus, resulted in a change on the gap spacing, as shown in Fig. 1. The cavity itself consists of a copper block with a cylindrical cavity of 9.94 mm diameter and 1.4 mm height, as well as a 1 mm diameter cylindrical copper post. The cavity design also consists of a coarse tuning mechanism, adjusted by placing spacers of various thicknesses between the structure A and B. In this way, it was possible to raise the pin and set the gap spacing as required for the particular experiment.

[a] Author to whom correspondence should be addressed. Electronic mail: natalia.docarmocarvalho@research.uwa.edu.au.

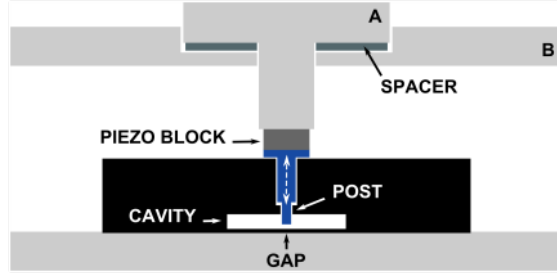

FIG. 1. (Color online). Illustration of the cross-section of the tunable cavity resonator. The cavity is placed into an enclosure consisting of two parallel plates; the distance between them is fixed. The cavity is screwed to the lower plate and the post is bonded to the PZT block, which is attached to the structure A. Structure A passes through the upper plate (structure B), which is lifted by spacers of different thicknesses to alter and set the initial gap spacing.

A number of experimental runs were undertaken at room temperature by adjusting different resonant frequencies using a variety of spacers. Then, for each configuration, fine-tuning was achieved by implementing the PZT actuator. A DC voltage was applied to the PZT block; consequently its excitation provided an additional decrease of the reentrant cavity gap proportionally to the applied voltage. Fig. 2 shows a diagram of the measurement set up. The reentrant cavity resonant modes were measured in transmission, S21, with a vector network analyzer (VNA), where the loop probes were adjusted to be in the under coupled regime and excited the transverse azimuthal magnetic field component oscillating around the post.

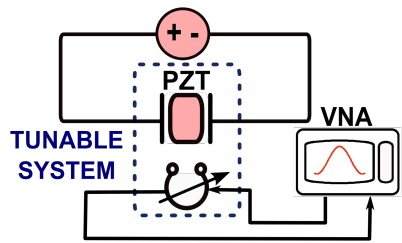

FIG. 2. (Color online). Schematic diagram of the measurement system for the tunable resonant cavity experiment. For cryogenic experimental runs, the tunable system was placed into a cryogen free cryocooler.

A tuning range from 2.8 GHz to 12.5 GHz was achieved with the mechanical tuning spacer mechanism. Fig. 3 shows a variety of experimental runs, where $f$ represents the resonant mode initial frequency as set by the spacer only, and $\mathit{\Delta f_{eff}}$ is the effective frequency shift of the cavity due the dynamic range of the PZT excitation, with $\mathit{\Delta U}$ from 0 to 60 V. A very large dynamic range of 1.1 GHz was achieved from an initial frequency of 5.7 GHz at room temperature. This was reduced to nearly 140 MHz at cryogenic temperatures, which indicates that the cryogenic efficiency of the PZT material was reduced by a factor of approximately 7.9.

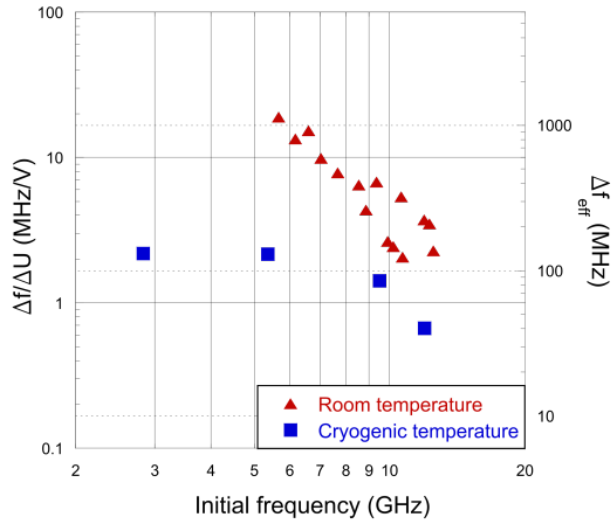

FIG. 3. (Color online). Frequency-tuning sensitivity $\mathit{\Delta f}/\mathit{\Delta U}$ . The dynamic range, $\mathit{\Delta f_{eff}}$, was caused by a voltage shift of 60 V. $\mathit{\Delta f}/\mathit{\Delta U}$ and $\mathit{\Delta f_{eff}}$ are presented as a function of the initial resonant frequency $f$ ($\mathit{\Delta U}$ = 0 V).

Following this, the frequency-tuning sensitivity of the cavity $\mathit{\Delta f}/\mathit{\Delta U}$ was estimated. It has been shown the resonant frequency decreases with smaller gap[8], introducing a large capacitance between the post and lid of the cavity. This high capacitance renders the device as a very sensitive displacement sensor. Hence, the gap changes induce a very large frequency shift, being responsible for the high tunability. Therefore, given that all measurements were approximately linear over the tuning range, Fig. 3 also illustrates the general trend of increased sensitivity of $\mathit{\Delta f}/\mathit{\Delta U}$, as $f$ is decreased. This value was determined by doing a series of five measurements, thus each point on Fig. 3 is verified and proven to be repeatable, with the average presented in the diagram.

For cryogenics runs, from 3 K to 5 K, it was observed the PZT sensitivity was less than the values at room temperature. Despite this, the technique allowed frequency shifts of up to 138.5

MHz. This was obtained at $f$ = 2.8 GHz, corresponding to a sensitivity of 2.2 MHz/V. At room temperature, the highest tuning sensitivity measured was 19 MHz/V at 5.7 GHz.

### III. PIEZOELECTRIC PROPERTIES

The high tunability and sensitivity of the reentrant cavity resonator make the device suitable for characterizing the sensing properties of the PZT actuator from room to cryogenic temperatures. Taking $x$ as the gap size, the $\Delta x/\Delta U$ parameter represents a figure of merit of the piezoelectric actuator, i.e., the displacement achieved per unit input voltage. At a set temperature, the $\Delta x/\Delta U$ is intrinsically related to the PZT properties and should be independent of the reentrant mode frequencies and initial gap setting. This parameter may be calculated by:

$$\frac{\Delta x}{\Delta U} = \frac{\Delta f}{\Delta U} \times \frac{\Delta x}{\Delta f} \qquad (1)$$

In the range we are calculating the gap spacing, the reentrant cavity lumped equivalent model[11] presents a very good agreement with measurements[26], then we calculated $\Delta f/\Delta x$ from the following equation:

$$f(x) = \sqrt{2 \pi h \mu_0 \ln\frac{r_{cav}}{r_{post}} \left( \frac{\pi r_{post}^2 \epsilon_0}{x} + 4 r_{post} \epsilon_0 \left( 1 + \ln\frac{\sqrt{h^2 + (r_{cav} - r_{post})^2}}{2x} \right) \right)} \qquad (2)$$

Where, $h$ and $r_{cav}$ are the cavity height and radius, $r_{post}$ is the post radius and $\mu_0$, $\epsilon_0$ are the permeability and the permittivity of free space, respectively.

Thus, by dividing the $\Delta f/\Delta U$ obtained experimentally and $\Delta f/\Delta x$ calculated from Equation 2, the value of $\Delta x/\Delta U$ may be determined. Fig. 4 shows $\Delta x/\Delta U$ for the different initial frequencies that was set for each run; these are negatives values, because a positive voltage causes the gap spacing to decrease as the voltage increases.

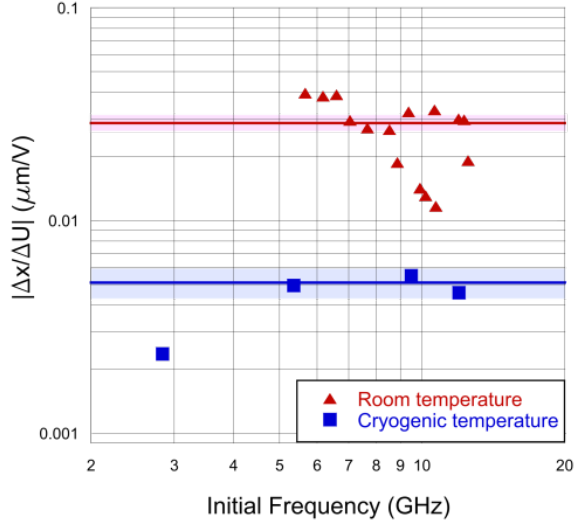

FIG. 4. (Color online). Modulus of the gap spacing displacement per unit volt, $|\Delta x/\Delta U|$, versus the initial reentrant resonant frequency. The solid lines are the average values of $|\Delta x/\Delta U|$, while the detached areas correspond to the standard error. The solid triangles and squares represent the determination by measurements, based on the data in Fig. 3.

It is expected that $\Delta x/\Delta U$ is constant at a specific temperature and in no way should depend on the gap size or the resonance frequency of the cavity. Thus, the precision of the $\Delta x/\Delta U$ determination at a constant temperature can be improved by calculating its average along with its standard error across all measurements. At room temperature, the average value is calculated to be -28.8 ± 3 nm/V and at 4K, -5.1 ± 1 nm/V. These results mean that the PZT actuator at cryogenic temperatures works at approximately 18% of its performance at room temperature, or is reduced in sensitivity by a factor of 5.7.

## IV.   CAVITY ELECTROMECHANICAL PROPERTIES

The cavity resonant mode was also dynamically tuned by the PZT actuator, which was driven by an AC voltage by using a Direct Digital Synthesizer (DDS). This induced a time varying gap displacement, and consequently produced a frequency modulation at the PZT excitation rate. The measurement setup used is shown in Fig. 5. The top part (A) describes the PZT excitation, and the bottom part (B) illustrates the readout system. The frequency synthesizer (SYNTH) excited the cavity resonant frequency and the signal was transmitted through the resonator. The reflected signal was mixed with the synthesizer frequency. In this configuration, the demodulated signal from the output of the mixer in the frequency discriminator is proportional to the mechanical displacement of the cavity post, which was measured using a Fast Fourier Transform (FFT) spectrum analyzer.

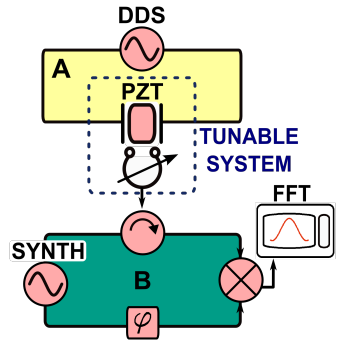

FIG. 5. (Color online). Schematic diagram of the measurement system for frequency modulation.

The modulation frequency was tuned from 2 kHz to above 600 kHz, with the output of the FFT presented in Fig. 6. The prominent peaks are due to the frequency matching of the modulation frequency with the mechanical modes of the cavity post, allowing readout of the mechanical resonances. Simulations based on Finite Element Method revealed two mechanical modes in this frequency range, at 150 kHz and 340 kHz. The corresponding mode structure is on the right of each peak in Fig. 6. The lower frequency represents the mode of vibration of the whole pin, while the upper mode represents the vibration of the top part of the pin, which has a smaller diameter. Thereby, we can say this tunable system presents itself as an excellent electromechanical sensor, proved to be efficient for dynamic characterization of mechanical resonances.

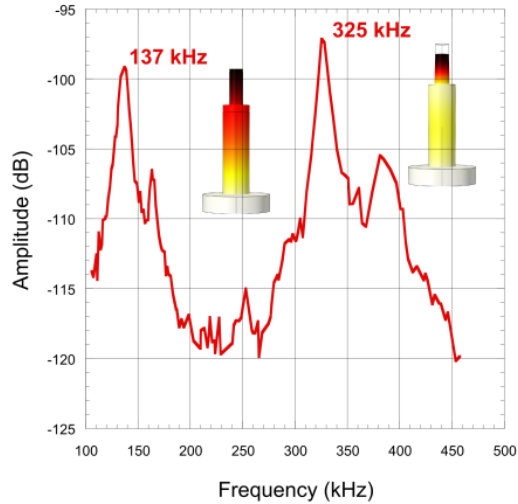

FIG. 6. (Color online). Mechanical modes of cavity pin obtained due the PZT dynamic excitation. Dark colors indicate maximum and light colors minimum displacement.

## V.     CONCLUSION

A highly tunable voltage coupled piezoelectric reentrant cavity was constructed and tested, which showed a great performance as a tunable cavity and an electromechanical sensor. It has been demonstrated the system was able to be fine-tuned electronically over a large frequency range: 1.1 GHz at room temperature (19%) and 138.5 MHz (5%) at cryogenic temperatures, at resonant frequencies of 5.7 GHz and 2.8 GHz, respectively. In addition, the cavity exhibited a high tuning sensitivity. At room temperature, a sensitivity of 2.3 MHz/V to 19 MHz/V was achieved, while at cryogenic temperatures the range was from 0.7 MHz/V to 2.2 MH/V. Moreover, we have shown that the voltage tunable reentrant cavity can be used for determining piezoelectric sensitivity at different temperatures, allowing a novel and highly precise technique for in situ characterization of piezoelectric materials. Also, we have illustrated it was possible to excite and detect the mechanical modes of the cavity through the piezoelectric excitation. Consequently, the piezoelectric coupled reentrant cavity can be used as a tool for investigating electromechanical systems.  Therefore, we can conclude that interesting applications can be drawn with such a device; in summary, with a single cavity it is possible to tune frequency resonances, characterize PZT actuators and have an electromechanical sensor.

ACKNOWLEDGMENTS

This research is supported by the Australian Research Council FL0992016, CE110001013 and by the Conselho Nacional de Desenvolvimento Científico e Tecnológico (CNPq – Brazil).